\documentclass[noshowpacs,twocolumn, amsmath, amssymb, aps, prx,superscriptaddress]{revtex4-2}

\usepackage{xcolor,graphicx}
\usepackage[colorlinks=true]{hyperref}
\usepackage{soul}
\usepackage{colortbl}
\usepackage{booktabs}
\usepackage[normalem]{ulem}
\usepackage[normalem]{ulem}

\usepackage{dcolumn}
\usepackage{bm}
\usepackage{comment}
\usepackage{nccmath,amsmath,lipsum}
\usepackage{svg}
\usepackage{siunitx} 
\usepackage{physics}
\let\qty\SI 
\usepackage{qcircuit}
\usepackage[mathlines]{lineno}

\newcommand{\cnn}{Universit\'e Paris-Saclay, Centre de Nanosciences et de Nanotechnologies, CNRS, 10 Boulevard Thomas Gobert, 91120, Palaiseau, France}

\newcommand{\UD}{University of Delaware, Newark, DE 19716}

\newcommand{\quandela}{Quandela SAS, 7 Rue L\'eonard de Vinci, 91300 Massy, France}

\newcommand{\ParisCite}{Universit\'e Paris Cit\'e, Centre de Nanosciences et de Nanotechnologies, CNRS, 10 Boulevard Thomas Gobert, 91120, Palaiseau, France}

\begin{document}
 
\title{Dynamical decoupling of a quantum dot spin in a micropillar cavity for spin-multiphoton entanglement}

\author{H.~Huet}
\thanks{These authors contributed equally to this work.}
\affiliation{\cnn}
\affiliation{\quandela}

\author{P.~R.~Ramesh}
\thanks{These authors contributed equally to this work.}
\altaffiliation{Present address: Johns Hopkins Applied Physics Laboratory, Laurel, Maryland 20723, USA}
\affiliation{\cnn}
\affiliation{\UD}

\author{R.~Frantzeskakis}
\affiliation{\quandela}

\author{L.~Couronn\'{e}}
\affiliation{\cnn}
\affiliation{\quandela}

\author{P.~Steindl}
\affiliation{\cnn}

\author{V.~Guichard}
\affiliation{\cnn}

\author{M.~Morassi}
\author{A.~Lemaître}
\author{I.~Sagnes}
\affiliation{\cnn}

\author{M.F.~Doty}
\affiliation{\UD}

\author{L.~Lanco}
\affiliation{\cnn}
\affiliation{\ParisCite}

\author{D.~A.~Fioretto}
\affiliation{\cnn}
\affiliation{\quandela}

\author{S.~C.~Wein}
\affiliation{\quandela}

\author{P.~Senellart}
\affiliation{\cnn}

\author{O.~Krebs}
\email{olivier.krebs@universite-paris-saclay.fr}
\affiliation{\cnn}

\date{\today}

\begin{abstract}

Graph states of mutually entangled photons are key resources for quantum computation and communication and can be generated by leveraging the entanglement between a single resident spin and emitted photons from a charged semiconductor quantum dot (QD). This approach is intrinsically limited by the decoherence of the spin. We study how to mitigate this decoherence with dynamical decoupling of an electron spin in the weak transverse magnetic field regime using spin echo and Carr-Purcell-Meiboom-Gill (CPMG) techniques. Application of these techniques allows us to extend the coherence time of a spin by more than two orders of magnitude, extracting a $T_2^{CPMG}$ of \qty{298\pm53}{\nano\second}. We further demonstrate that this technique is compatible with the generation of a spin-photon-photon entangled state at a high rate enabled by a micropillar cavity, with a 20\% improvement in simulated state fidelity when using dynamical decoupling. These results pave the way for generating larger and more complex entangled states with QDs.

\end{abstract}

\maketitle

\section{\label{intro}Introduction}

Entangled photonic graph states are fundamental building blocks in the progression towards scalable fault-tolerant quantum computing ~\cite{Raussendorf2001,Raussendorf2007,Briegel2009} and large-scale quantum networking~\cite{Azuma2023,Kimble2008}. Semiconductor quantum dots (QDs) confining a single trapped charge have emerged as a promising platform for producing such entangled states. By leveraging the optical selection rules connecting the two spin ground states to the excited trion states, the deterministic generation of spin-multiphoton entangled states has recently been demonstrated~\cite{Coste2023,Cogan2023,Su2024,Huet2025}. A primary obstacle to scaling these resource states is the limited coherence time of the spin, which is typically dominated by the inhomogeneous dephasing time ($T_2^*$) of the resident (central) spin arising from the hyperfine interaction with the nuclear spin bath~\cite{Urbaszek2013}. The $T_2^*$ for an electron spin in an InGaAs QD is typically only $\sim$\qty{2}{ns}~\cite{Bechtold2015,Stockill2016}. Due to its reduced hyperfine interaction~\cite{Fischer2008,Eble2009,Cogan2018}, the spin of the valence-band heavy hole is less sensitive to dephasing, but the coherence time is still limited to a few tens of nanoseconds~\cite{Coste2023a,Huthmacher2018}.\\
\indent Several techniques exist for extending the coherence time of the spin. For example, dynamical decoupling techniques such as spin echo use optical pulses to coherently rotate the spin during its precession, allowing it to rephase by the end of the precession cycle~\cite{Hahn1950,Press2010,degreve2011}. This effectively mitigates the time-averaged dephasing of the spin arising from the slowly varying Overhauser field, bringing the coherence time closer to its intrinsic $T_2$ time~\cite{Urbaszek2013}. Spin echo has been shown to extend the measured coherence time $T_2^{SE}$ of an electron spin in an InGaAs QD into the microsecond regime~\cite{Stockill2016,Zaporski2023}. Another method for mitigating dephasing due to the interaction between the central spin and the nuclei is to narrow the nuclear spin distribution~\cite{Sun2012}, a technique sometimes referred to as nuclear spin cooling. Such techniques have been demonstrated with both an electron~\cite{Ethier-Majcher2017,Gangloff2019} and a hole spin~\cite{Prechtel2016,Hogg2024}, achieving $T_2^*$ times in the tens of microseconds.\\
\indent A combination of dynamical decoupling and nuclear spin cooling techniques was recently demonstrated within a spin-photon entanglement protocol~\cite{Meng2024}. However, using nuclear spin cooling requires a strong ($>$\qty{1}{\tesla}) magnetic field where only time-bin encoding of the entangled states is possible~\cite{Lee2019}. To generate entanglement between the spin and photon polarization (the Lindner and Rudolph scheme~\cite{Lindner2009}), one must operate in the weak ($<$\qty{100}{\milli\tesla}) magnetic field regime in which the spin state is controlled via coherent Larmor precession about the transverse magnetic field. This method offers several advantages for rate and scalability and has produced the best results to date for graph state generation with QDs~\cite{Huet2025,Su2024}.\\ 
\indent In this work, we explore the use of dynamical decoupling to extend the coherence time in a weak magnetic field ($<\qty{100}{\milli\tesla}$) regime. Using a bright, high-purity single photon source based on an InGaAs QD-micropillar cavity device~\cite{Somaschi2016}, we demonstrate over one hundred-fold improvement in the coherence time of the resident electron spin by using all-optical dynamical decoupling. We then demonstrate the generation of a spin-photon-photon entangled state where the addition of spin echo improves the state purity and simulated entanglement fidelity.

\section{\label{se}Ultrafast spin echo}

\begin{figure}
\includegraphics[width=\linewidth]{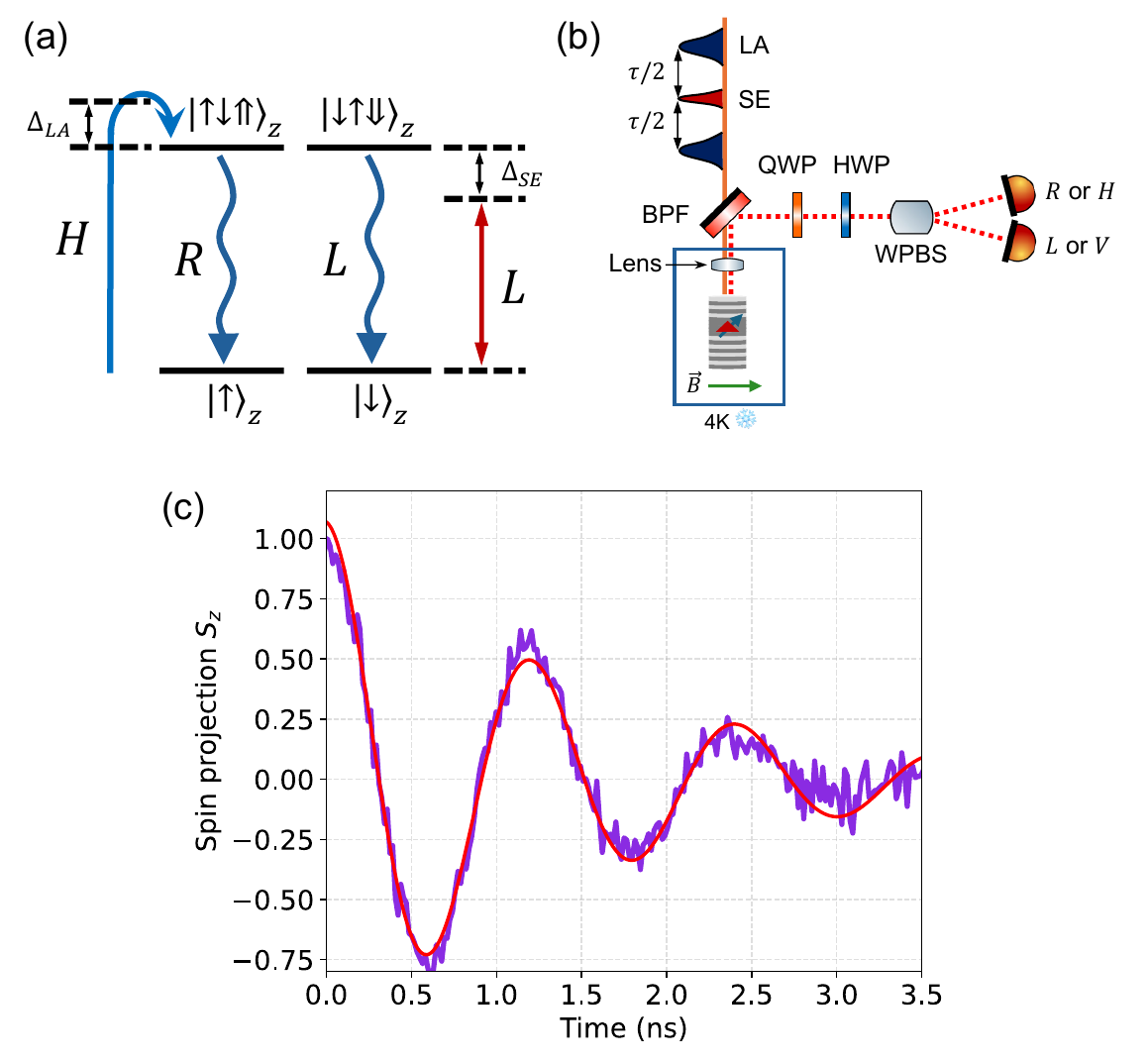}
\caption{(a) Optical selection rules of a negatively charged QD in the $z$-basis. Blue-detuned phonon-assisted excitation is linearly ($H$) polarized while corresponding optical transitions are circularly $R/L$ polarized. Red-detuned spin echo is circularly $L$ polarized. (b) Simplified depiction of the experimental setup for measuring spin coherence using photon polarization. BPF = narrow band-pass filter, QWP = quarter wave plate, HWP = half wave plate, WPBS = Wollaston polarizing beam splitter. A simplified laser pulse sequence of excitation (LA) and spin echo (SE) is shown. Superconducting nanowire single photon detectors (SNSPDs) measure single photons in polarization bases $R/L$ or $H/V$. (c) Measure of the normalized projection $S_z$ of the electron spin under free precession in a \qty{100}{\milli\tesla} transverse magnetic field, revealing a $T_2^*\approx\qty{1.5}{\nano\second}$. The red line is a damped sinusoidal fit to the data.}
\label{fig:device}
\end{figure}

We use an InGaAs QD hosting a single trapped electron precisely positioned in the center of a micropillar cavity device~\cite{Dousse2008,Somaschi2016}. The device is electrically contacted to deterministically control the QD charge state~\cite{Nowak2014}. The device is held at $T=\qty{4}{\kelvin}$ within an optical cryostat that includes a superconducting magnetic coil that applies a \qty{60}{\milli\tesla} transverse magnetic field. The optical selection rules for the negative trion $\ket{X^-}$ state are depicted in Fig.~\ref{fig:device}a. These selection rules perfectly map the spin state of the electron to the circular polarization ($R$ or $L$) of the emitted single photons, thus providing an avenue to initialize and readout our spin state and generate spin-photon entanglement. The QD is addressed using quasi-resonant longitudinal acoustic (LA) phonon-assisted excitation, which preserves the optical selection rules~\cite{Thomas2021,Coste2023a}. The single photons emitted from the QD are separated from the excitation path with three narrow band-pass filters (BPF), after which they are sent to a polarization tomography setup consisting of waveplates,  Wollaston prisms, and superconducting nanowire single photon detectors (SNSPDs).

\indent We first measure the free-induction decay of the spin without spin echo to estimate the $T^*_2$ time. This is achieved using continuous-wave excitation with linear polarization~\cite{Coste2023a,Serov2024}. The measurement begins with a first single photon detected in the circularly right $R$ polarization state, effectively heralding the spin in the $\ket{\uparrow}_z$ state (abbreviated as $\ket{\uparrow}$ for the remainder of this work). After a time delay $t$, a second emitted photon is detected in the opposite polarization state ($L$), serving as the readout of the spin state in a \textit{pump-probe}-like experiment. From this measurement, the normalized spin projection along the $z$-axis is extracted as $S_z(t)=\frac{I_R(t)-I_L(t)}{I_R(t)+I_L(t)}$, where $I_{R}(t)$ and $I_{L}(t)$ are the conditional detection counts in the $R$ and $L$ basis at time $t$, respectively. The time evolution of the measured $S_z$ is shown in Fig.~\ref{fig:device}c, where the envelope capturing the decay of the free precession oscillations provides an estimate of $T_2^*\approx\qty{1.5}{\nano\second}$.

\begin{figure*}[t]
\includegraphics[width=0.9\linewidth]{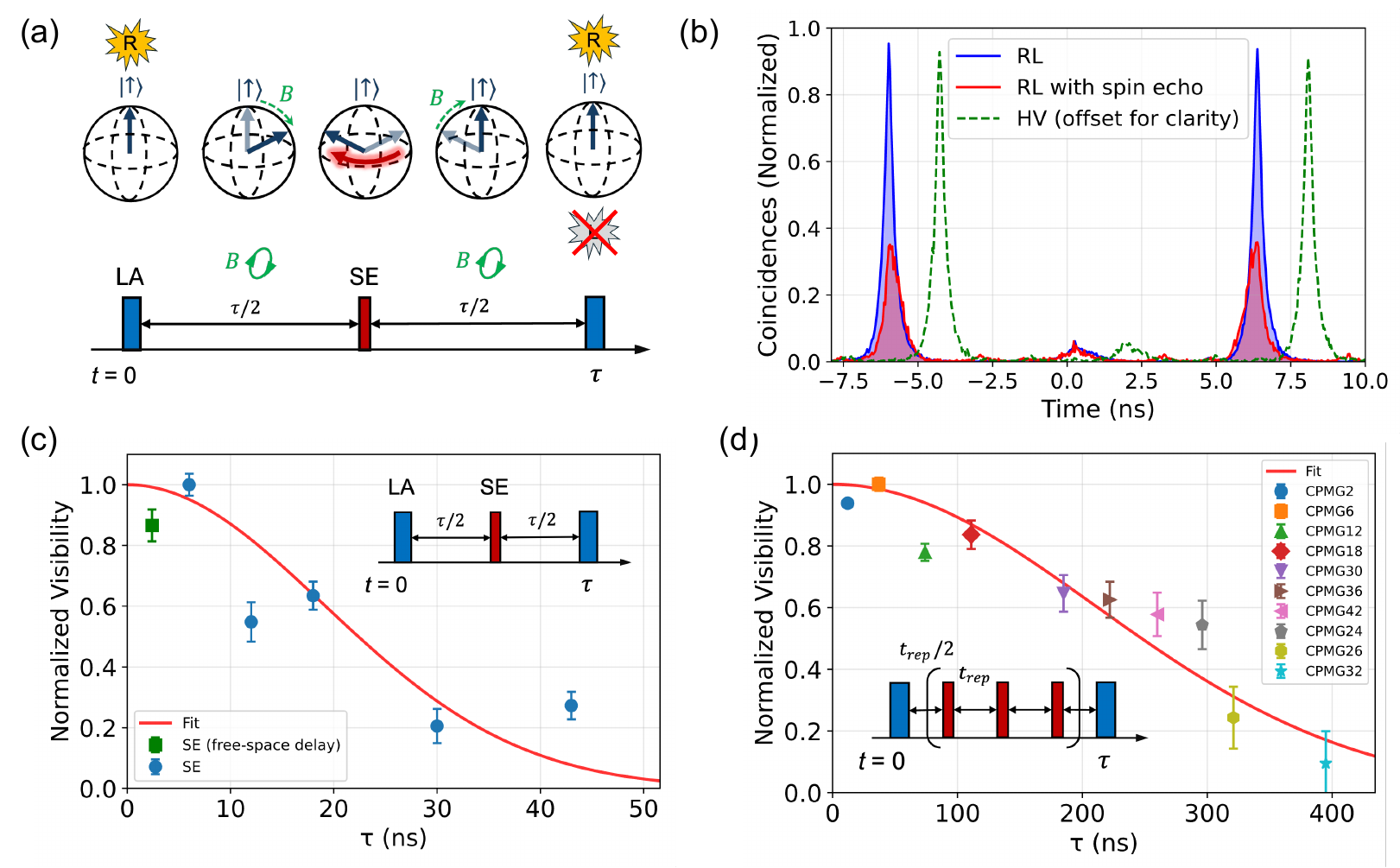}
\caption{(a) Schematic of the spin echo pulse sequence and corresponding evolution of the spin state in the Bloch sphere, beginning with detection of an $R$ photon to herald the spin in state $\ket{\uparrow}$. If the spin echo is successful, the second and final photon should be detected in $R$ (click shown in yellow), or conversely an absence of detection in $L$ (click crossed out). (b) Histogram of two-photon cross-correlation for $\tau=\qty{6}{\nano\second}$, with the measurement polarization basis of the first and last photon indicated in the legend. Histogram for $H/V$ is offset in time for clarity. (c-d) Coherence time measurements of normalized visibility $\mathcal{V}$ versus time delay $\tau$ with (c) spin echo and (d) CPMG dynamical decoupling. Legend in (d) indicates the number of refocusing pulses $N$ for the given data point. Respective pulse sequences are shown in the insets of the plots. Red lines represent Gaussian decay fits for each respective case. Error bars represent one standard deviation.}
\label{fig:t2}
\end{figure*}

To enhance the spin coherence time using the spin echo (SE) technique, we use the experimental pulse sequence shown in Fig.~\ref{fig:device}b. Specifically, we add an ultrafast (\qty{4}{\pico\second}), red-detuned, circularly-polarized pulse that coherently rotates the spin about its $z$-axis (the optical and growth axis) without leading to photon emission~\cite{Press2008,Greilich2009,Godden2012,Stockill2016}. The LA excitation and spin echo pulses are prepared from the same femtosecond laser by using pulse-shaping and spectral filtering to spatially separate the excitation and SE pulses. This separation enables independent control over polarization, power, and tailored pulse sequences using electro-optic modulators (EOMs). A schematic of the experimental setup is shown in Fig.~\ref{fig:setup} in the Supplemental Material~\cite{supp}. The excitation pulse is blue-detuned from resonance by $\Delta_{LA}=-\qty{0.7}{\nano\metre}$, while the spin echo (SE) pulse has a red detuning of $\Delta_{SE}=+\qty{1.2}{\nano\metre}$. \\
\indent The pulsed excitation is provided by a laser with a repetition rate of $\qty{81}{\mega\hertz}$, corresponding to a pulse every $\qty{12.35}{\nano\second}$. The laser repetition rate is then doubled using a free-space delay with retro-reflectors, such that we obtain one pulse every $t_{rep}\simeq\qty{6}{\nano\second}$. To probe time scales longer than $t_{rep}$, we use the EOMs to selectively suppress a variable number of pulses $N$, which allows us to access different delays $\tau$:
\begin{equation}
    \label{eq:reprate}
    \tau=(N+1)t_{rep}
\end{equation}
For example, suppressing two out of every three pulses ($N=2$) yields $\tau\simeq\qty{18}{\nano\second}$, while suppressing six out of every seven pulses gives $\tau\simeq\qty{42}{\nano\second}$. Note that the same pulse sequence is always used for both LA-phonon assisted excitation pulses and spin echo pulses, but the SE pulses are delayed by a time $\tau/2$, ensuring they arrive exactly between two successive photon emissions as shown in Fig.~\ref{fig:device}b.\\
\indent The measurement of spin coherence with the addition of a spin echo pulse starts with the heralding of the spin state via detection of a photon with $R$ circular polarization. The spin then undergoes free precession around the magnetic field axis for time $\tau/2$, after which the SE pulse performs a $\pi$-rotation about the optical axis. After the spin precesses for an additional $\tau/2$, a second excitation pulse is applied and a second photon is measured. In the absence of decoherence, the $\pi$-rotation induced by the spin echo pulse will cause the electron spin to be in the state $\ket{\uparrow}$ at time $\tau$, which means that the second emitted photon can only have $R$ polarization. A schematic illustrating this measurement is shown in Fig.~\ref{fig:t2}a, where we see the evolution of the electron spin in the Bloch sphere and the corresponding photon detection events.\\
\indent To quantify how SE improves $T_2^*$ we measure the time-resolved cross-correlations between the first and second photons in the $R$/$L$ basis as a function of the time delay $\tau$ between these photons. For each value of $\tau$, we measure the probability (visibility) of events in which the first photon is detected in $R$ and the second photon is detected in $L$. An example of the correlation histogram used to compute this visibility for $\tau=\qty{6}{\nano\second}$ is shown in Fig.~\ref{fig:t2}b. We first note that counts at $t=0$ time delay are strongly suppressed due to single photon anti-bunching. The correlations at $t=\pm\tau$ then provide information about the spin state via the measured polarization of the second photon. If an applied SE pulse perfectly mitigates spin dephasing, then no $L$ photons would be detected from the second photon emission, meaning a complete suppression of the $R$/$L$ correlations at delay $\tau$. The amplitude of these first-delay cross-correlations (shown in red in Fig.~\ref{fig:t2}b) thus provides a measure of the spin coherence.\\ 
\indent To plot this visibility as a function of $\tau$ we normalize the data to account for the photon collection probability. If the spin had fully decohered within the time $\tau$, we would expect no suppression of the $R$/$L$ correlations at delay $\tau$. This behavior is observed in the correlation histogram obtained \textit{without} use of a spin echo pulse (blue) in Fig.~\ref{fig:t2}b. Here the spin can be considered to be nearly completely decohered because a time delay of $\tau= \qty{6}{\nano\second}$ is four times the intrinsic $T_2^*\approx\qty{1.5}{\nano\second}$. Similarly, if we measure two-photon correlations in the $H$/$V$ polarization basis (dashed green lines in Fig.~\ref{fig:t2}b) the spin is not heralded, which means it is in a mixed state equivalent to the case of complete decoherence. We can see that the $H/V$ histogram in Fig.~\ref{fig:device}b matches exactly the case without spin echo. We therefore use the $H$/$V$ histogram as our reference and define a visibility metric $\mathcal{V}$ based on the integrated area of correlations ($A_{X/X}$) at time $\tau$ to quantify the degree of spin coherence:
\begin{equation}\label{eq:visibility}
    \mathcal{V} = 1 - \frac{A_{R/L}}{A_{H/V}}
\end{equation}
\indent To evaluate the spin echo coherence time $T_2^{SE}$, we simply repeat this measurement for a series of time delays, using the EOMs to adjust the delay $\tau$ between successive laser pulses. The normalized results ($\mathcal{V}_\tau/\mathcal{V}_{\text{max}}$) are shown by the blue data points in Fig.~\ref{fig:t2}c. The solid red line represents a Gaussian decay fit, of the form $y = e^{-(x/T_2)^2}$, to all of the experimental data~\cite{Stockill2016,Zaporski2023}. From this fit, we can extract a coherence time of $T_2^{SE}\simeq\qty{27\pm4}{\nano\second}$, which is more than ten-times longer than the natural $T_2^*$ without spin echo. We note that Fig.~\ref{fig:t2}c includes one data point (green square) generated using free-space delay lines, rather than EOMs, to set a time delay $\tau=\qty{2.4}{\nano\second}$ that is shorter than $t_{rep}$. This is also the time delay used below in the entanglement generation experiment. Including this data point validates that the spin echo technique implemented via free space delays works on the time scale needed for entanglement generation.\\
\indent We next explore the generalization of spin echo to multiple pulses, which is known as the Carr-Purcell-Meiboom-Gill (CPMG) technique~\cite{Carr1954,Meiboom1958,Cywinski2008}. To do so we vary both the time delay $\tau$ and the number of equally-spaced refocusing pulses between the two LA excitation pulses, as shown in the inset of Fig.~\ref{fig:t2}d. The number of refocusing pulses ($N$) is determined by the time delay and the repetition rate of the laser as explained in Eq.~\ref{eq:reprate}. For example, when the repetition rate is $R_{rep}\simeq\qty{162}{\mega\hertz}$, a delay of $\tau=\qty{260}{\nano\second}$ results in 42 refocusing pulses (``CPMG42'' in Fig.~\ref{fig:t2}d), whereas for $R_{rep}\simeq\qty{81}{\mega\hertz}$, $\tau=\qty{395}{\nano\second}$ corresponds to only 32 refocusing pulses (``CPMG32'' in Fig.~\ref{fig:t2}d). As shown in Fig.~\ref{fig:t2}d, the use of CPMG pulse sequences dramatically extends the spin coherence. We again fit the data with a Gaussian as shown by the solid red line. From this we extract a coherence time of $T_2^{CPMG}\simeq\qty{298\pm53}{\nano\second}$, which is over one hundred times greater than the natural $T_2^*$ of the electron spin.\\ 
\indent In the weak magnetic field regime, the quadrupolar coupling of nuclear spins to the electric field gradients associated with inhomogeneous strain and chemical alloying dominates over the nuclear Zeeman interaction and induces fluctuations of the Overhauser field at frequencies of order 10 MHz~\cite{Sinitsyn2012,Hackmann2015,Stockill2016,Bulutay2012,chekhovich2012structural}. While this limits the capacity to use spin echo techniques to further extend spin decoherence times, the $T_2$ times achieved here, on the order of hundreds of nanoseconds, represent the current state-of-the-art for an InGaAs QD in a weak magnetic field. Below we show that this extension of the $T_2$ time enables significant improvements to spin-photon entanglement generation.

\section{\label{entanglement}Spin-multiphoton entanglement}

\begin{figure*}
\includegraphics[scale=0.55]{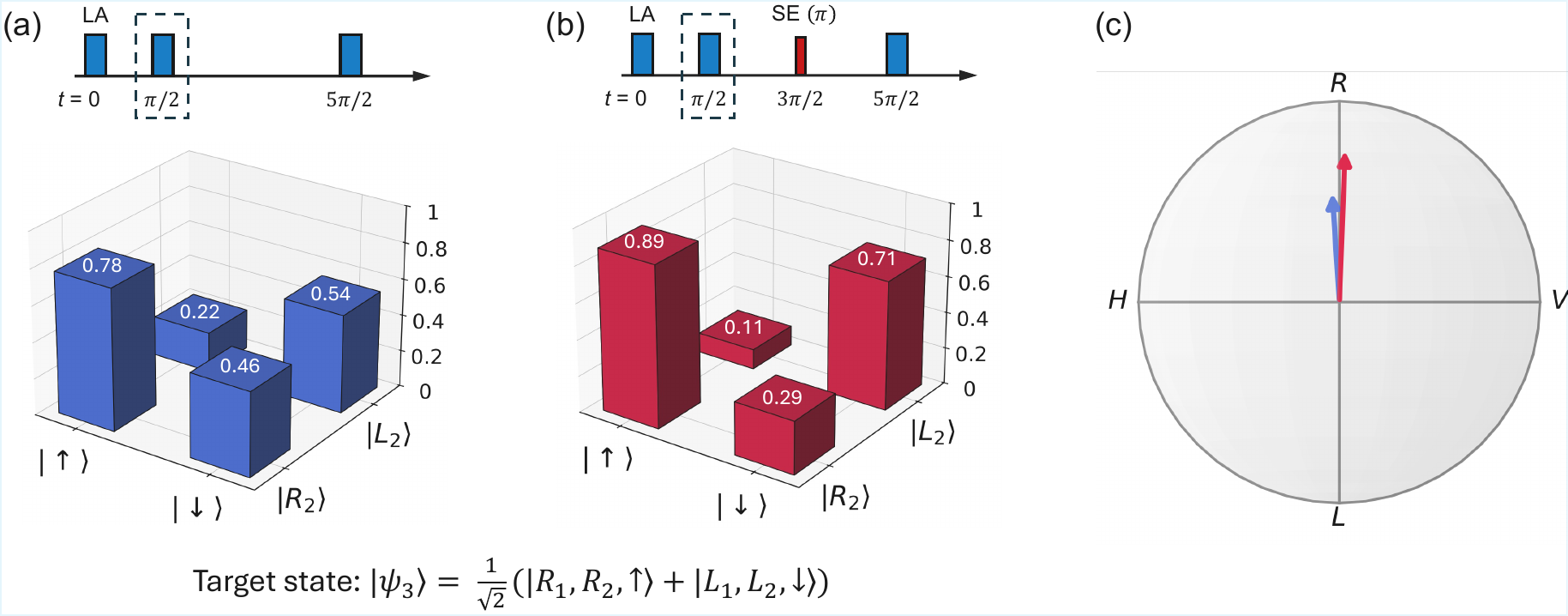}
\caption{(a,b) Truth tables for a spin-photon-photon GHZ state showing conditional probabilities of measuring the second photon in $R$ or $L$ given that the spin state is projected onto $\ket{\uparrow}$ or $\ket{\downarrow}$ via the measurement of the third photon. Panel (a) represents the case with free spin evolution, while panel (b) includes a spin echo pulse applied between photons \#2 and \#3. Corresponding pulse sequences are shown above. (c) Stokes vectors of photon \#2 conditioned on the spin being projected onto the $\ket{\uparrow}$ state, shown for the case without spin echo (blue) and with spin echo (red), corresponding to the truth tables in (a) and (b) respectively. The purities of the represented polarization states are $0.61 \pm 0.05$ without spin echo and $0.81\pm0.05$ with spin echo.}
\label{fig:ghz}
\end{figure*}

We now evaluate the compatibility of the spin echo technique with deterministic spin-photon-photon entanglement generation. We use the Lindner and Rudolph protocol, as implemented in Ref.~\cite{Coste2023}, which synchronizes excitation pulses to the coherent Larmor precession of the electron spin under a \qty{60}{\milli\tesla} transverse magnetic field to trigger the sequential emission of photons whose polarization states are entangled with the spin state~\cite{Lindner2009}. In this case, at $t=0$ the spin is heralded in state $\ket{\uparrow}$ by the detection of a photon in $\ket{R}$. The spin is then allowed to precess for a period of $t_{12}=\qty{600}{\pico\second}$, corresponding to a $\pi/2$ precession of the spin, before it is again excited to create the entangled state $\frac{1}{\sqrt{2}}\left(\ket{R_2}\ket{\uparrow}+\ket{L_2}\ket{\downarrow}\right)$. Each spin component of this superposition continues to precess for $t_{23}=\qty{2400}{\pico\second}$, a full $2\pi$ precession, before the third and final photon is emitted, producing the spin-photon-photon Greenberger-Horne-Zeilinger (GHZ) state $\ket{\psi_3}=~\frac{1}{\sqrt{2}}\left(\ket{R_2R_3}\ket{\uparrow}+\ket{L_2L_3}\ket{\downarrow}\right)$~\cite{GHZ}. Note that in this description of the protocol we neglect the finite lifetime of the trion $T_1^{ph}$, which is taken into account in the simulations reported below because it directly impacts the entanglement fidelity~\cite{Ramesh2025}.

\indent The $\ket{\psi_3}$ state is characterized by measuring photon \#3 in the $R/L$ polarization basis (effectively implementing a spin readout and thus disentangling the spin), and performing a full polarization tomography of photon \#2. From these measurements, we construct truth tables of the conditional probabilities between the polarization state of photon \#2 and the spin state as measured via the polarization of photon \#3. In Fig.~\ref{fig:ghz}a-b we compare the truth tables characteristic for GHZ state generation without spin echo (a) and with spin echo (b) between second and third photon generation. In the truth tables we see evidence of the $\ket{\psi_3}$ state, with positive correlations between $\ket{\uparrow}$ and $\ket{R_2}$ and similarly between $\ket{\downarrow}$ and $\ket{L_2}$. The use of a spin echo pulse significantly enhances the strength of these correlations, as shown in Fig~\ref{fig:ghz}b. These results can also be represented by the Stokes vector of the conditional state of photon \#2 calculated from the detected counts in all three polarization bases ($R/L$, $H/V$, $D/A$). Fig.~\ref{fig:ghz}c depicts the Stokes vector on the Poincar\'{e} sphere. From the length of this Stokes vector we quantify the state purity when the spin was measured in $\ket{\uparrow}$. Without spin echo, the state purity has a value of $0.61 \pm 0.05$. With spin echo, the state purity is improved to $0.81 \pm 0.05$. Both Stokes vectors are slightly tilted from the target state $R$ due to spin precession in the excited state during the $T_1^{ph}=\qty{200}{ps}$ trion emission lifetime~\cite{Ramesh2025}, which could be mitigated by using a device with a shorter emission lifetime. While the truth tables and Stokes vectors are not direct evidence of entanglement, they are required for the successful generation of the target 3-partite GHZ state, and can be used to simulate the expected entangled state fidelities.

We therefore use a model to simulate the results of the three-qubit GHZ state generation with and without spin echo. The model we use~\cite{Coste2023,Huet2025} captures spin dephasing via a time-dependent Overhauser field due to the quadrupolar interaction coupling~\cite{Sinitsyn2012}. The results of the simulations (Fig.~\ref{fig:truthtable}~\cite{supp}) show excellent agreement with the experimental data presented in Fig.~\ref{fig:ghz}. From these simulations we extract three-qubit GHZ state fidelities $F_{3-GHZ}=0.63$ and $F_{3-GHZ}^{SE}=\nolinebreak0.76$, indicating a significant improvement in state fidelity when dynamical decoupling is used. A complete list of parameters used for the simulation can be found in Supplementary Table~\ref{table: model parameters}~\cite{supp}. 

\section{\label{conclusion}Conclusion \& Perspectives}

We explored the use of optical dynamical decoupling techniques to extend the electron spin coherence time in a QD-cavity device. This is the first such demonstration that is compatible with the generation of spin-photon entangled states in a weak transverse magnetic field, achievable with cryogenic permanent magnets~\cite{Steindl2023}. We employed a two-photon measurement technique wherein the detection of the first photon heralds the initialization of the spin state, and the second photon acts as a readout of the spin state after a time $\tau$. Using this method, we found that the use of a spin echo pulse at $t=\tau/2$ optically rephases the spin and extends the coherence time from $T_2\simeq$\qty{1.5}{\nano\second} to $T_2^{SE}\simeq$\qty{20}{\nano\second}, which represents a tenfold improvement. By extending the spin echo to multiple pulses (CPMG), we further extend the coherence time to $T_2^{CPMG} \simeq$\qty{300}{\nano\second}, a hundredfold improvement. Finally, we demonstrate that the addition of spin echo significantly improves the quality of the spin-photon-photon target state, with a $\sim$20\% improvement in simulated state fidelity. We note that these 3-partite entangled states can be generated at a high rate~\cite{Coste2023}, which is important for enabling further scalability.\\
\indent The results of this work take an essential step towards fault-tolerant quantum computing using photonic graph states or spin-photon entangled states. For example, it has been recently shown~\cite{Wein2024a} that photonic caterpillar graph states~\cite{Hilaire2023,Pettersson2025} generated with a QD source using a resident spin~\cite{Huet2025} can be used to reach fault-tolerant computation while maintaining a low resource overhead with as few as 14 entangled photons. Such generation could be realized using recently demonstrated protocols~\cite{Huet2025} with $T_2$ times on the order of \qty{20}-\qty{30}{\nano\second} such as we have achieved here. Similarly, a recently proposed hybrid spin-optical computing architecture based on a network of emitters generating spin-photon entangled states~\cite{DeGliniasty2024,Hilaire2023} exhibits a fault-tolerant threshold of $\sim2\%$ of the gate time $t_{gate}$ to the coherence time $T_2$. For photon-photon gates times around 10-\qty{100}{\nano\second}, this would require coherence times in the hundreds of nanoseconds to a few microseconds. As we show here, the CPMG pulse sequence can extend the spin coherence time into this feasibility window. Moreover, the fast entanglement rates enabled by this protocol are critical for repeat-until-success gates used in both the photonic graph state and spin-optical architectures~\cite{Wein2024a,DeGliniasty2024}. The results reported here thus pave the way for extending the coherence time in a high-rate spin-photon entanglement protocol into a regime that will enable scalable, fault-tolerant architectures. 

\begin{acknowledgements}
    The authors wish to acknowledge Noah Shofer and Nathan Coste for their useful inputs. This work was partially supported by the Paris Ile-de-France Région in the framework of DIM SIRTEQ, the European Union’s Horizon 2020 FET OPEN project QLUSTER (Grant ID 862035), Horizon CL4 program under the grant agreement 101135288 for EPIQUE project, by the European Commission as part of the EIC accelerator program under the grant agreement 190188855 for SEPOQC project, the Plan France 2030 through the projects ANR22-PETQ-0011, ANR-22-PETQ-0006 and ANR-22-PETQ-0013, the French National Research Agency (ANR) project SPIQE (ANR\-14\-CE32\-0012), and a public grant overseen by the French National Research Agency as part of the``Investissements d’Avenir'' programme (Labex NanoSaclay, reference: ANR\-10\-LABX\-0035). P.R.R. acknowledges the financial support of the Fulbright-Universit\'{e} Paris-Saclay Doctoral Research Award and the UNIDEL Distinguished Graduate Fellowship. M.F.D. acknowledges support from the National Science Foundation (2217786). This work was done within the C2N micro nanotechnologies platforms and partly supported by the RENATECH network and the General Council of Essonne.
\end{acknowledgements}

\bibliography{Biblio}

\clearpage 

\begin{widetext}
\section*{Supplementary information}
\setcounter{figure}{0}
\renewcommand{\thefigure}{S\arabic{figure}}

\subsection{Extended data and visibility calculation}
The data in Fig.~\ref{fig:t2} uses the peak areas from correlation histograms in only the $R/L$ and $H/V$ bases for various time delays $\tau$, accessed using an EOM to pick certain pulses from repetition rate of the laser. For the spin echo (SE) data, the ratio of the time delay $\tau=2-\qty{45}{\nano\second}$ to the temporal window of single photon detection $t_{window}\approx\qty{300}{\pico\second}$ is quite small, so the double exponential profile of the correlation peaks can be well-resolved. In this regime, the data for the first-delay correlation peaks can be fit accurately with such a Laplace distribution or double exponential function of the form:

\begin{equation}\label{eq:laplace}
    f(x)=ae^{-\abs{x-c}/w}
\end{equation}

where $a$ is the amplitude, $c$ is the center, and $w$ is the width. Then, an area can be calculated analytically as $A=2\times a\times w$. This has an accompanying uncertainty

\begin{equation}
    \delta A_{exp} = 2w \cdot \delta a = 2w\sqrt{a}
\end{equation}

which arises from Poisson counting statistics. The amplitude $a$ represents a photon count which has a Poisson uncertainty of $\delta a = \sqrt{a}$.

For the CPMG data where $\tau\gg t_{window}$, the double exponential fit becomes challenging, and instead trapezoidal numerical integration is used. However, knowing that the ideal function is still a double exponential, uncertainty can be calculated with this reference.

\begin{equation}
    \delta A_{trapz} = \left|\frac{x_{int}^3}{12N^2} \cdot f''(x)\right|
\end{equation}

Here, $x_{int}$ is the width of the integration window, $N$ is the number of integration intervals, and $f''(\xi)$ is the second derivative of the double exponential function is Equation~\ref{eq:laplace} with its essential parameters $a,w$ estimated from the data within the integration window. 

The peak area error from either of these methods ($\delta A_{exp}$ for spin echo data and $\delta A_{trapz}$ for CPMG data) can then be propagated into the uncertainty for the visibility as shown below.

\begin{equation}
    \delta \mathcal{V} = \sqrt{\left(\frac{\delta A_{RL}}{A_{HV}}\right)^2 + \left(\frac{A_{RL} \cdot \delta A_{HV}}{A_{HV}^2}\right)^2}
\end{equation}

This is reflected in the error bars in Fig.~\ref{fig:t2}. 

\subsection{Simulation methods}

The simulation of spin-photon entanglement is based on a four-level trion system that evolves following a Markovian master equation to describe the impact of spontaneous emission. The hyperfine interaction between the electron spin and the nuclei is captured by an additional Zeeman Hamiltonian to model the fluctuating Overhauser (OH) field with an isotropic Gaussian distribution. 
The Hamiltonian of the four-level system is described as follows:
\begin{equation}
    H=\frac{\Delta_e}{2}\sigma^{(e)}_y+\frac{\Delta_h}{2}\sigma^{(h)}_y
\end{equation}
where $\sigma^{(e)}_y=i(\ket{\downarrow}\bra{\uparrow}-\ket{\uparrow}\bra{\downarrow})$ and $\sigma^{(h)}_y=i(\ket{\downarrow\uparrow\Downarrow}\bra{\downarrow\uparrow\Uparrow}-\ket{\downarrow\uparrow\Uparrow}\bra{\downarrow\uparrow\Downarrow})$ are the electron and hole spin operators, and $\Delta_{e}, \Delta_h=\mu_Bg_eB,\mu_Bg_hB$ are the Zeeman splitting energies due to a static transverse magnetic field $B$ with $\mu_B$ as the Bohr magneton and $g_e$ ($g_h$) as the effective electron (hole) $g$ factor. 

To capture the hyperfine interaction between the electron spin and nuclei we average the electron spin dynamics over a Gaussian distribution of Overhauser field fluctuations ($B_{OH}$) impacting the electronic state through an additional Zeeman-like term in the Hamiltonian.
\begin{equation}\label{eq:ohfield}
    H_{OH}=\frac{1}{2}g_e\mu_B\vec{B_{OH}}\cdot\vec{\sigma}^{(e)}
\end{equation}
\indent Here $\vec{\sigma}^{(e)}=(\sigma^{(e)}_x,\sigma^{(e)}_y,\sigma^{(e)}_z)$ and the $\vec{B_{OH}}=(B_{x,OH},B_{y,OH},B_{z,OH})$ can capture anisotropic effects. To approximate the quadrupolar interactions with the nuclei arising from strain effects, we use a normalized vector $B_{\hat{n},qr}$ corresponding to the direction of an effective  magnetic field $B_{Q}$ about which a given instance of the Overhauser field rotates at the frequency $\omega_Q/2\pi\approx 15$~MHz~\cite{Sinitsyn2012}. The resulting time-dependent Overhauser field is then used in place of the static Overhauser field Hamiltonian with all other simulation elements being the same as Ref.~\cite{Huet2025}.

\indent The impact of spontaneous emission is described using a Markovian master equation of the form:
\begin{equation}\label{eq:lindbladall}
    \frac{d}{dt}\rho(t)=-i[H+H_{O},\rho(t)]+\gamma D_{\sigma_R}\rho(t)+\gamma D_{\sigma_L}\rho(t)
\end{equation}
where $\gamma=1/T_1^{ph}$ of the photon, $D\rho=\sigma\rho\sigma^{\dagger} -\sigma^{\dagger}\sigma\rho/2 -\rho\sigma^{\dagger}\sigma/2$ is the action of the Lindblad dissipator superoperator, and $\sigma_R = \ket{\uparrow}\bra{\downarrow\uparrow\Uparrow}$ $\left(\sigma_L=\ket{\downarrow}\bra{\downarrow\uparrow\Downarrow}\right)$ is the optical lowering operator coupled to right (left) circularly-polarized light.

The instantaneous spin rotation pulses (SE pulses) can be modelled as:
\begin{equation}
    R_{SE}(\theta)=exp(-i\theta\sigma^{(e)}_z/2)
\end{equation}
and a dephasing channel (a Liouvillian superoperator) affects the electron after the SE prior to the next precession. For more details on the simulation of optical spin rotations, see the Supplementary Information of \cite{Huet2025}.

\begin{figure*}[h!]
 \includegraphics[scale=0.55]{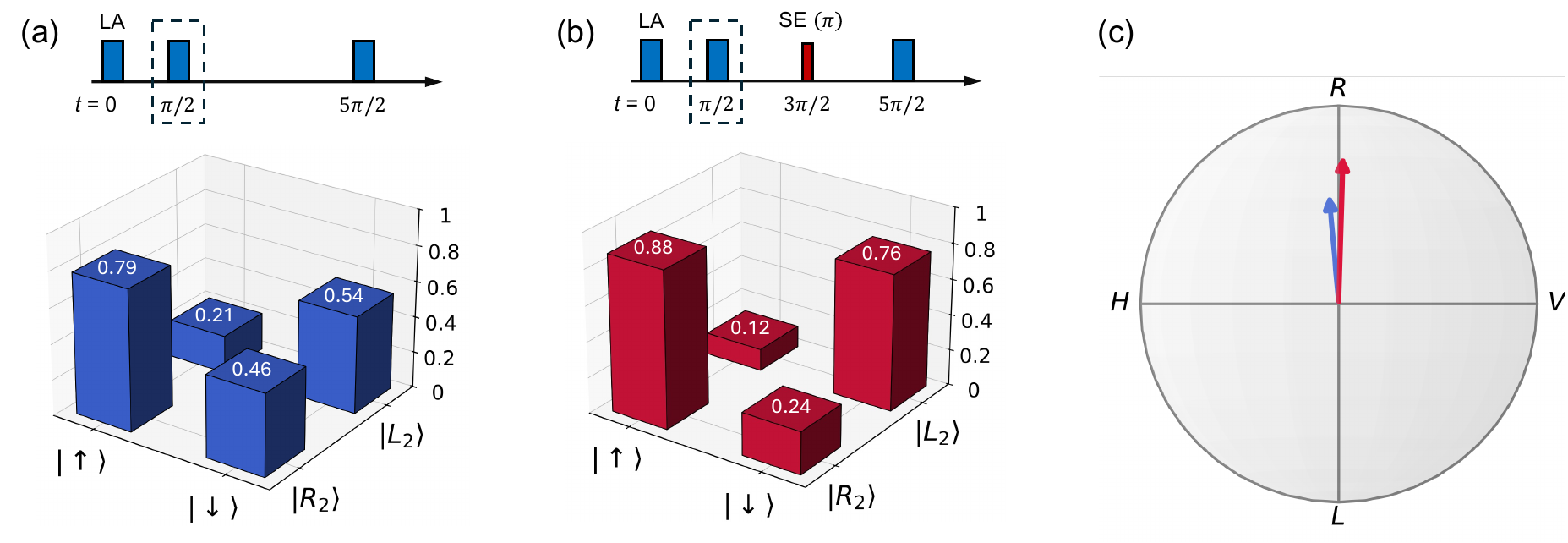}
\caption{Simulation of spin-photon-photon entanglement without and with spin echo. Simulated conditional probabilities of measuring
the second photon in R or L when the spin state is up or down without spin echo (a) and with spin echo (b). (c) Stokes
vectors of photon $\#$2 without spin echo (blue) and with spin echo (red), with their respective purities. The purities of the photon polarization states are $0.53$ and 
and $0.81$, respectively.}
\label{fig:truthtable}
\end{figure*}

Finally, we simulate the generation of a spin-photon-photon GHZ state, replicating the experiment shown in the main text. We use the aforementioned master equation methods to simulate each step of the entanglement process, including the spontaneous emission of a photon, optical spin rotation, and the spin dephasing from the Overhauser and quadrupolar fields. The parameters of the simulation are shown in Table~\ref{table: model parameters}. The radiative lifetime $T_1^{ph}$, external magnetic field amplitude $B_{ext}$, Overhauser field distribution $B_{OH}$ and electron and hole $g$-factors ($g_e$ and $g_h$) were determined in previous works~\cite{Coste2023,Huet2025} and are therefore kept fixed. The quadrupolar-induced precession frequency $\omega_Q/2\pi$ and excitation polarization angle $\theta_{exc}$, which defines the excitation laser polarization state $e^{-i\theta_{exc}}\ket{R}+e^{i\theta_{exc}}\ket{L}$, are adjustable parameters tuned to accurately model the experimental results.

From this simulation we extract the conditional probabilities of measuring the second photon in $R$ or $L$ when the spin state is $\ket{\uparrow}$ or $\ket{\downarrow}$.  The resulting simulated truth tables and Stokes vectors are shown in Supplementary Fig.~\ref{fig:truthtable}. 

\begin{table}[!ht]
\centering
\rowcolors{2}{gray!12}{white}
\begin{tabular}{ccc}
  \toprule
  \textbf{Symbol} & \textbf{Description} & \textbf{Value} \\
  \midrule
  $T_1^{ph}$ & Radiative lifetime & 200 ps \\
  $\theta_{exc}$ & Excitation polarization & $0.4\pi$ \\
  $B_{OH}$ & Effective Overhauser field distribution & 9 mT\\
  $\omega_Q/2\pi$ & Quadrupolar-induced precession frequency & 15 MHz\\
  $B_{ext}$ & External magnetic field amplitude & 60 mT\\
  $g_e$ & Electron $g$-factor & 0.6\\
  $g_h$ & Hole $g$-factor & 0.3\\

  \bottomrule
\end{tabular}
\caption{Model parameters to simulate the conditional probabilities of the second photon for the spin-photon-photon experiments.}
\label{table: model parameters}
\end{table}

The simulated truth tables and Stokes vectors match the experimental results quite well. Similarly to the experiment, we observe here approximately a $17\%$ improvement for the purity and $20\%$ in the fidelity when using spin echo.

\subsection{Experimental setup}

A schematic of the experimental setup used in this work is shown in Fig.~\ref{fig:setup}. Optical pulses originate from a single femtosecond laser with a \qty{12.35}{\nano\second} repetition rate. They are shaped using a $4f$-line and a spatial light modulator. The pulse repetition rate is doubled using a retro-reflector, then separated spectrally into the longitudinal-acoustic phonon-assisted excitation pulses (LA) and the spin echo pulses (SE) using a narrow band pass filter (BPF). The optical path for each of the two pulses contains an electro-optic modulator (EOM) to pick pulses at different time delays. There is a neutral density filter (NDF) to control the power of each pulse. A combination of waveplates, polarizers, and fiber paddles serve as a polarization controller (PC) to set any arbitrary polarization in each path. The pulses recombine with another BPF and are sent into a optical fiber. For the entanglement generation, the EOMs are not used, so the recombined pulses go through a free-space delay line to adjust the time delays before being sent via fiber to the QD-cavity device contained in an optical cryostat kept at 4K. A superconducting magnet inside the cryostat applies an in-plane magnetic field. The single photons emitted from the QD-cavity device are separated from the incoming laser pulses by another BPF and then sent to a polarization tomography setup. Each tomography path consists of a HWP, QWP, a Wollaston polarizing beam splitter, and a two superconducting nanowire single photon detectors (SNSPD). For the spin coherence measurement, only one path is used, set to $R/L$ polarization bases. For the entanglement measurement, three paths are needed: two in $R/L$ for initialization and readout, and one that measures in all bases for the full tomography of photon \#2.

\begin{figure}[h!]
\includegraphics[width=\linewidth]{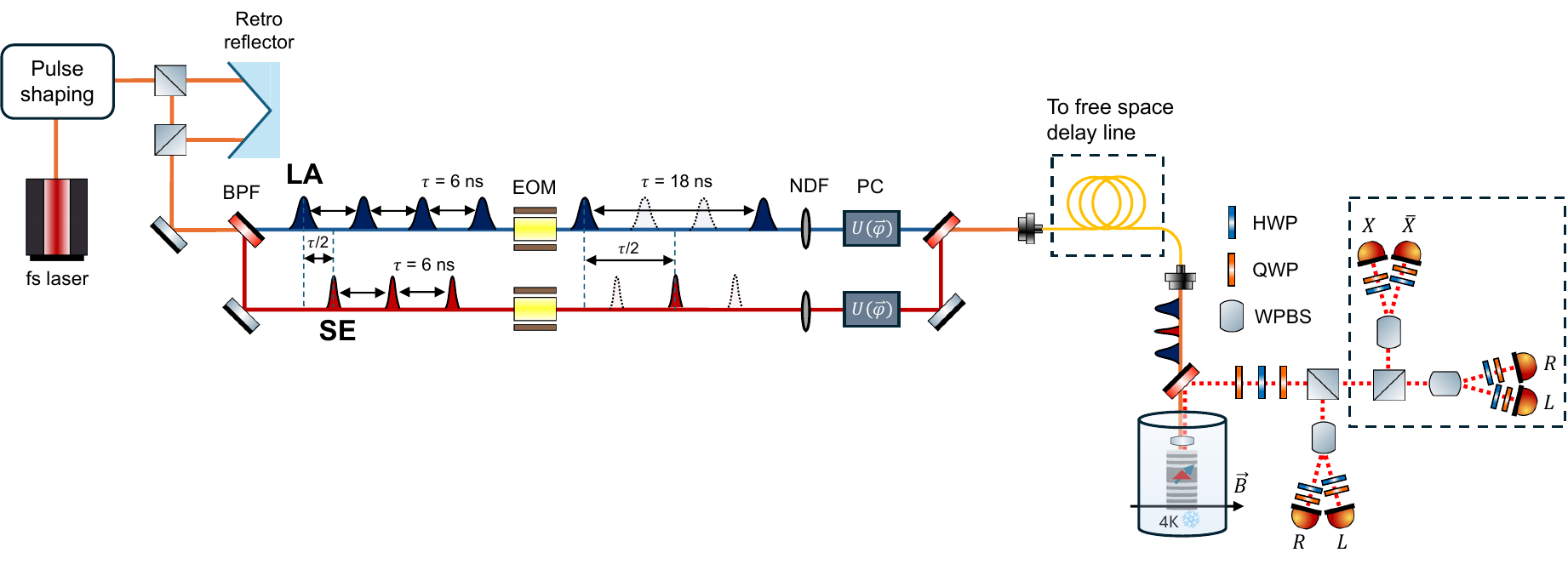}
\caption{Schematic of the experimental setup. Dashed boxes indicate the parts of the setup used only for the entanglement-generation experiment. LA and SE denote longitudinal-acoustic phonon-assisted excitation and spin echo pulses, respectively. Electro-optic modulators (EOM) are used to suppress pulses and generate tailored pulse sequences with controlled delays. Neutral density filters (NDF) control the optical power, while polarization optics include half-wave plates (HWP), quarter-wave plates (QWP), and Wollaston polarizing beam splitters (WPBS).} 
\label{fig:setup}
\end{figure}

\end{widetext}
\end{document}